\documentclass[letterpaper, 10 pt, onecolumn, conference]{ieeeconf}  

\IEEEoverridecommandlockouts                              
\usepackage{graphicx} 
\usepackage{caption}
\usepackage{subfigure}
\usepackage{amsmath} 
\usepackage{amssymb}  
\usepackage[ruled]{algorithm2e}

\renewcommand\baselinestretch{2.0}

\title{\LARGE \bf
Simultaneous Greedy Analysis Pursuit for Compressive Sensing of Multi-Channel ECG Signals
}

\author{Yurrit Avonds, Yipeng Liu*, and Sabine Van Huffel 
\thanks{This work was supported by Research Council KUL: GOA MaNet, PFV/10/002 (OPTEC),  IDO 08/013 Autism, several PhD/postdoc and fellow grants; Flemish Government: FWO: PhD/postdoc grants, projects:  FWO G.0427.10N (Integrated EEG-fMRI), G.0108.11 (Compressed Sensing) G.0869.12N (Tumor imaging); IWT: TBM070713-Accelero, TBM080658-MRI (EEG-fMRI), PhD Grants; iMinds; Belgian Federal Science Policy Office: IUAP P6/04 (DYSCO, `Dynamical systems, control and optimization', 2007-2011); ESA AO-PGPF-01, PRODEX (CardioControl) C4000103224
EU:  RECAP 209G within INTERREG IVB NWE programme, EU HIP Trial FP7-HEALTH/ 2007-2013 (n$бу$ 260777) ( Neuromath (COST-BM0601). \emph{Asterisk indicates corresponding author.}}%
\thanks{All the authors are with ESAT-STADIUS / iMinds Future Health Department, Dept. of Electrical Engineering, KU Leuven, Kasteelpark Arenberg 10, box 2446, 3001 Leuven, Belgium.
        {\tt\small 	email: yurrit.avonds@student.kuleuven.be; $ \{ $yipeng.liu;sabine.vanhuffel$ \}$@esat.kuleuven.be }}%
}

\begin{document}

\maketitle
\thispagestyle{empty}
\pagestyle{empty}

\begin{abstract}

This paper addresses compressive sensing for multi-channel ECG. Compared to the traditional sparse signal recovery approach which decomposes the signal into the product of a dictionary and a sparse vector, the recently developed cosparse approach exploits sparsity of the product of an analysis matrix and the original signal. We apply the cosparse Greedy Analysis Pursuit (GAP) algorithm for compressive sensing of ECG signals. Moreover, to reduce processing time, classical signal-channel GAP is generalized to the multi-channel GAP algorithm, which simultaneously reconstructs multiple signals with similar support. Numerical experiments show that the proposed method outperforms the classical sparse multi-channel greedy algorithms in terms of accuracy and the single-channel cosparse approach in terms of processing speed.

\end{abstract}



\section{Introduction}

Mobile ECG monitors allow patients to be continuously monitored, even when they are not in a hospital. They have limited available energy, which limits the amount of data they can transmit. Compression can be used to reduce the amount of transmitted data over these energy-hungry wireless body sensor network links. Currently, the signal is sampled at a considerably high sampling rate and then compressed, which requires temporary storage of a large amount of uncompressed data before reducing it.

Compressive sensing (CS) provides a solution by compressing while sampling, avoiding the high sampling rate and storage requirements imposed by the Nyquist rate. For illustration convenience, the sampling model is formulated in the discrete setting as $\mathbf{y} = \pmb{\phi} \mathbf{x}$, where  $ \pmb{\phi} \in {\mathbb{R}^{n \times N}} $ and $n < N$, making measurement $\textbf{y}$ a compressed form of signal $\textbf{x}$.

To exploit sparsity for signal recovery, a signal $\textbf{x}$ can be represented as $\mathbf{x} = \pmb{\psi} \mathbf{s}$, i.e. a linear combination of the columns from a dictionary or basis $\pmb{\psi} \in {\mathbb{R}^{n \times M}} $. If $\textbf{s}$ contains a small number of nonzero elements, it is said to be a sparse vector. To estimate the sparse vector $\textbf{s}$ from $\textbf{y}$, $\pmb{\phi}$ and $\pmb{\psi}$, we can formulate the synthesis-based $\ell_0$-minimization problem:
\begin{equation}
	\underset{\mathbf{s}}{min.} ||\mathbf{s}||_0,~ \: s.~t.~~ \: \mathbf{y} = \pmb{\phi} \pmb{\psi} \mathbf{s}
	\label{eq:sparseL1minimization}
\end{equation}
where $ ||\mathbf{s}||_0 $ counts the number of nonzero elements in \textbf{s}.

Recently, a cosparse approach was introduced, stating that an analysis matrix $\pmb{\Omega} \in {\mathbb{R}^{p \times N}} $ can provide a sparse representation $\pmb{\Omega x}$. \cite{nam2013cosparse} In order to recover $\textbf{x}$, an analysis-based $\ell_0$-minimization problem can be formulated:
\begin{equation}
	\underset{\mathbf{x}}{min.} ||\pmb{\Omega} \mathbf{x}||_0,~~ \: s.~t.~~ \: \mathbf{y} = \pmb{\phi} \mathbf{x}
	\label{eq:AnalysisSparseL1minimization}
\end{equation}


To solve the cosparse approach based nonconvex programming problem (\ref{eq:AnalysisSparseL1minimization}), there are two popular ways. The Analysis-by-Synthesis (ABS) method is a convex relaxation that rewrites the problem in the form of a synthesis-based problem \cite{cleju2012analysis}. Greedy Analysis Pursuit (GAP) is a greedy algorithm that is the analysis duality of Orthogonal Matching Pursuit (OMP) \cite{nam2013cosparse}. Other algorithms include Analysis IHT (AIHT), analysis HTP (AHTP), analysis CoSaMP (ACoSaMP) and Analysis SP (ASP). The four greedy-like algorithms are developed corresponding to their synthesis counterparts in an analysis setting \cite{giryes2013greedy, nam2011cosparse}.

Compared to previous works on CS for single-channel (SC) ECG \cite{liu2013multi, mamaghanian2011compressed}, the processed data is large-scale, since multiple channels are processed simultaneously. Assuming channels share a similar support \cite{kim2012compressive}, multi-channel (MC) algorithms can be used to recover the MC ECG signals with reduced computational complexity. To the best of our knowledge, all current MC methods, such as simultaneous orthogonal matching pursuit (SOMP) \cite{tropp2006algorithms}, the multiple response extension of the standard Sparse Bayesian Learning (M-SBL) \cite{wipf2006bayesian}, and Reduce MMV and Boost (ReMBO) \cite{mishali2008reduce}, are based on a sparse representation of the synthesis form. However, our research shows that the cosparsity based convex relaxation gives higher reconstruction accuracy for ECG signals than the classical sparse convex relaxation methods \cite{liu2013multi}.

In this paper, the SC GAP is generalized to process MC signals to reduce the total processing time for all channels. The algorithm is used for simultaneous reconstruction of MC ECG signals. The choice of a greedy algorithm is motivated by its lower computational complexity which is an important advantage for wireless ECG monitoring. Numerical experiments show that the proposed algorithm improves reconstruction accuracy without increasing computational complexity compared to sparse MC ECG signal reconstruction and reconstructs multiple ECG channels faster than the SC GAP algorithm without reducing the reconstruction accuracy.

The remaining part of the paper is organized as follows. In Section~\ref{sec:methods} the new reconstruction algorithm is presented. In Section~\ref{sec:experiments} the performed experiments are described. In Section~\ref{sec:results} the results of these experiments are presented. Finally, in Section~\ref{sec:conclusion} a conclusion about the results is drawn.

\section{Methods}
\label{sec:methods}

In this section, we first improve GAP by introducing some precalculations. Next, the Simultaneous Greedy Analysis Pursuit (SGAP) method is introduced.

\subsection{GAP}
\label{sub:GAP}

While OMP searches the support of the sparse vector $\textbf{s}$ to give a sparse estimate \cite{tropp2010computational}, the GAP algorithm finds the co-support (collection of zero elements) of $\pmb{\Omega} \mathbf{x}$. This is done by first initialising a full co-support $\hat{\Lambda}_k = \lbrace 1,2,3, ... ,p \rbrace$. In each iteration, the elements in the co-support that correspond to large values in the product of $\pmb{\Omega}$ and the current estimate of $\hat{\textbf{x}}$ are removed from the co-support estimate. Ideally, this leads to the removal of all nonzero elements from the co-support estimate \cite{nam2013cosparse}.

During the co-support update, only the index of the largest element in $\pmb{\alpha} = \Omega \hat{\mathbf{x}}_{k-1}$ is selected and removed from the co-support. It is also possible to select the indices corresponding to the $t$ largest values in $\pmb{\alpha}$. In this paper, $t$ is set to 10. The maximum number of iterations $K_{max}$ is set as $\lfloor{(p - t)/t}\rfloor$.

The second stopping criterion checks whether the norm of the solution estimate relative to the previous estimate is larger than in the previous iteration, since this indicates a decrease in reconstruction accuracy.

In \cite{nam2013cosparse}, it is stated that
\begin{equation}
	\hat{\mathbf{x}_k} =
	\begin{bmatrix}
		\pmb{\phi} \\
		\sqrt{\lambda} \pmb{\Omega}_{\hat{\Lambda}_k}
	\end{bmatrix}^\dagger
	\begin{bmatrix}
		\mathbf{y} \\
		\mathbf{0}
	\end{bmatrix} =
	(\pmb{\phi}^T \pmb{\phi} + \lambda \mathbf{\Omega}_{\hat{\Lambda}_k}^T \mathbf{\Omega}_{\hat{\Lambda}_k})^{-1} \pmb{\phi}^T \mathbf{y}
\label{eq:pseudoinverse}
\end{equation}

\noindent with $\lambda$ a small positive constant, set to $\lambda=0.05$. Since neither $\pmb{\phi}$ nor $\mathbf{y}$ change their value during the reconstruction process, the result of two matrix multiplications ($\pmb{\phi}^T \pmb{\phi}$ and $\pmb{\phi}^T \mathbf{y}$) can be precalculated. This results in a shorter processing time, especially when $\pmb{\phi}$ is large.

\subsection{SGAP}

SGAP extends GAP for MC measurements, assuming that the ECG signals in all channels share the same co-support, since each channel can be seen as a projection of the cardiac activity towards the electrode associated with the channel \cite{sameni2006multi}. All MC signals $\mathbf{x}$ and measurements $\mathbf{y}$ are now represented by capital letters ($\mathbf{X}$, $\mathbf{Y}$) to indicate that they are matrices instead of vectors.

To generalize the GAP, several steps should be adapted. The first difference is the addition of precalculations for speed improvement, as discussed in section~\ref{sub:GAP}. Secondly, since $\pmb{\alpha} = \Omega \hat{\mathbf{X}}$ is now a matrix, a different approach to find positions of nonzero elements in the co-support is required. A row-wise summation of $\pmb{\alpha}$ creates a vector with large values at positions that correspond to elements that are not part of the co-support of some or all of the channels \cite{tropp2006algorithms}. Finally, the second stopping criterion is adapted. Using a column-wise norm calculation, a vector with one value representing the change in norm compared to the previous solution estimate for each channel can be obtained. The algorithm iterates until the residual for at least one of the current channel approximations becomes larger than in the previous iteration.

Like the GAP algorithm, 10 elements are removed from the co-support in each iteration. $K_{max}$ is set like in the GAP algorithm as well. The complete SGAP algorithm is presented as Algorithm \ref{alg:sgap}.

\begin{algorithm}[htbp]
\small
\caption{Simultaneous Greedy Analysis Pursuit}
\label{alg:sgap}

$\bullet$ \textbf{In:} $\mathbf{Y} \in {\mathbb{R}^{n \times c}}$, $\pmb{\phi} \in {\mathbb{R}^{n \times N}}$, $\pmb{\Omega} \in {\mathbb{R}^{p \times N}}$

$\bullet$ \textbf{Out:} $\hat{\mathbf{X}} \in {\mathbb{R}^{N \times c}} $

$\bullet$ Initial Co-Support: $\hat{\Lambda}_k = \lbrace 1,2,3, ... ,p \rbrace$

$\bullet$ Initial Solution: $\hat{\mathbf{X}}_k = (\pmb{\phi}_{\phi} + \lambda \mathbf{\Omega}^T \mathbf{\Omega})^{-1} \pmb{\phi}_Y$

$\bullet$ Precalculations $\pmb{\phi}_{\phi} = \pmb{\phi}^T \pmb{\phi}$ ; $\pmb{\phi}_Y = \pmb{\phi}^T \mathbf{Y}$

\Repeat{ $k \geq K_{max}$ or $\mathbf{r}_k(i) > \mathbf{r}_{k-1}(i)$ for at least one $i \in \{ 1, ..., c \} $}
{
	$\bullet$ $k := k +1$

	$\bullet$ Analysis Representation: $\pmb{\alpha} = \Omega \hat{\mathbf{X}}_{k-1}$

	$\bullet$ Row-Wise Summation: $\pmb{\alpha}_R = \pmb{\alpha} .
	\begin{bmatrix}
	1 & 1 & \hdots & 1
	\end{bmatrix}^T $

	$\bullet$ Update Co-Support: $\hat{\Lambda}_k = \hat{\Lambda}_{k-1} \backslash \lbrace \underset{i \in \hat{\Lambda}_{k-1}}{argmax} \: |{\pmb{\alpha}_R}_i|\rbrace$

	$\bullet$ Update Solution: $\hat{\mathbf{X}}_k =
	(\pmb{\phi}_{\phi} + \lambda \mathbf{\Omega}_{\hat{\Lambda}_k}^T \mathbf{\Omega}_{\hat{\Lambda}_k})^{-1} \pmb{\phi}_Y$, where $\mathbf{\Omega}_{\hat{\Lambda}_k}$ is $\mathbf{\Omega}$ with rows not in $\hat{\Lambda}_k$ set to 0
}

where $\mathbf{r}_k(i) = 1 - \frac{||\hat{\mathbf{X}}_k(i)||_2}{||\hat{\mathbf{X}}_{k-1}(i)||_2}$ for $i \in \{ 1, ..., c \} $ and $\mathbf{X}_k(i)$ is the $i$-th column in the solution $\mathbf{X}$ of the $k$-th iteration.
	
	

\end{algorithm}

\section{Numerical Experiments}
\label{sec:experiments}

In the experiments, the MIT-BIH Arrhythmia Database \cite{moody2001impact,PhysioNet}, a commonly used ECG database, is used. It consists of 2-channel ECG recordings from 48 patients, sampled at 360 Hz. A total of 943 2-second segments of the first 23 patients in the database (the first 41 segments of each patient) are compressed and subsequently reconstructed. Segments of only 2 seconds are used to keep the processing time per segment relatively low.

In order to compare the performance of SGAP to a corresponding sparse algorithm, SOMMP - a faster version of SOMP where multiple dictionary elements are selected in each iteration - with a Daubechies wavelet dictionary is used. Moreover, separate channel reconstructions by OMMP and GAP are used to demonstrate the shorter processing times of MC algorithms. Both SOMMP and OMMP select 4 support elements per iteration, since it was empirically found that this renders the most accurate reconstructions for the dataset.

Different from the one in \cite{liu2013multi}, the analysis matrix used in these experiments is a second order derivative matrix, obtained by multiplying 2 first order derivative matrices. This type of matrix renders faster and more qualitative results than a first order derivative matrix.
\begin{equation}
	\mathbf{\Omega}_2 = 	\mathbf{\Omega}_1^2 =
	\left\lbrack \begin{smallmatrix}
	1 & -1 & 0 & 0 & \hdots & 0 \\
	0 & 1 & -1 & 0 & \hdots & 0 \\
	\vdots &  & \ddots & & & \vdots \\
	\vdots & & & \ddots & & 0 \\
	\vdots & & &  & 1 & -1\\
	0 & \hdots & \hdots & 0 & 0 & 1 \\
	\end{smallmatrix} \right\rbrack^2
\end{equation}

The amount of applied compression is expressed as the compression ratio $CR=n/N$. Values of CR ranging from 0.9 to 0.2 in steps of 0.1 are used in the experiments. In order to quantify the reconstruction accuracy, the percentage root-mean-square difference (PRD) was used \cite{mamaghanian2011compressed}.
\begin{equation}
	PRD = {||x - \hat{x}||_2}/{||x||_2}
	\label{eq:prd}
\end{equation}
To quantify the speed of the algorithm, the processing time until convergence is measured. For OMMP and GAP, these are calculated as the respective sums of the processing times from the separate channel reconstructions.

All experiments were performed in MATLAB\textregistered 2013a, on a quadcore 3.10GHz Intel\textregistered Core\texttrademark i5-3450 system with 8GB RAM, running CentOS 6.4 with Linux kernel version 2.6.32.

\section{Results}
\label{sec:results}

All of the results are presented as boxplots with five characteristic values. These are, in increasing order of magnitude: lower whisker ($LOW$), 25th percentile ($P25$), median ($MED$), 75th percentile ($P75$) and higher whisker ($HIGH$), where $LOW = P25 - w.(P75 - P25)$ and $HIGH = P75 + w.(P75 - P25)$, with $w = 1.5$. Outliers (red crosses) are values outside the $[LOW,HIGH]$ range.

\begin{figure}[ht]
\centering
\subfigure[]{
	\includegraphics[width = 3cm]{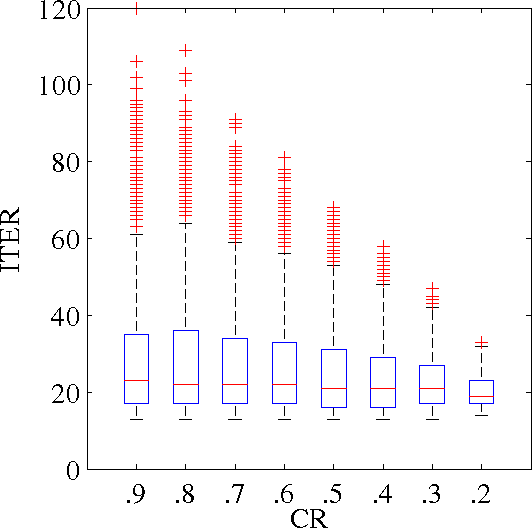}
    \label{subfig:sommpITER}
}
\subfigure[]{
	\includegraphics[width = 3cm]{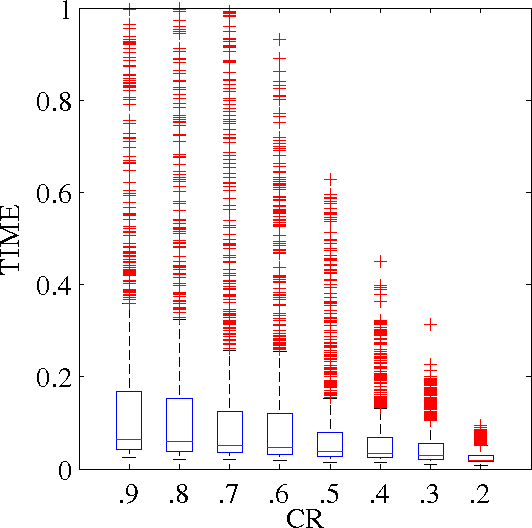}
    \label{subfig:sommpTIME}
}
\subfigure[]{
	\includegraphics[width = 3cm]{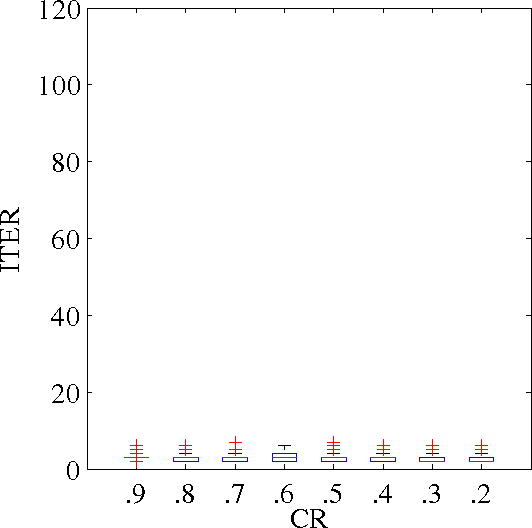}
    \label{subfig:sgapITER}
}
\subfigure[]{
	\includegraphics[width = 3cm]{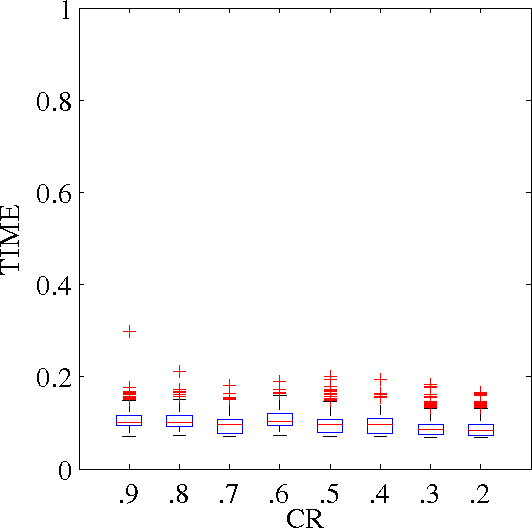}
    \label{subfig:sgapTIME}
}
\subfigure[]{
	\includegraphics[width = 3cm]{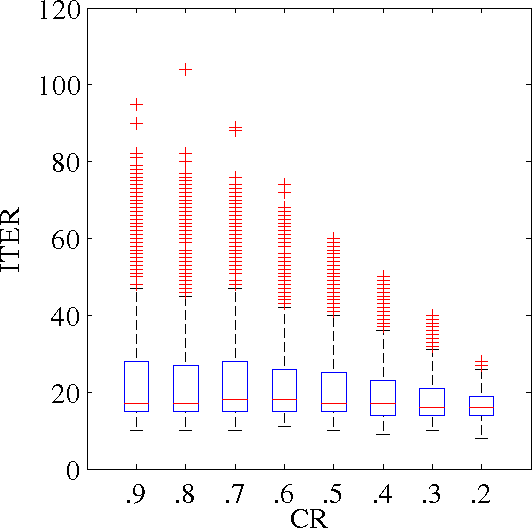}
    \label{subfig:ommpITER}
}
\subfigure[]{
	\includegraphics[width = 3cm]{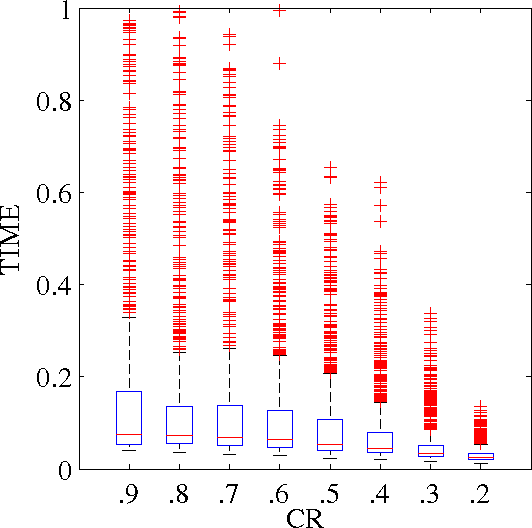}
    \label{subfig:ommpTIME}
}
\subfigure[]{
	\includegraphics[width = 3cm]{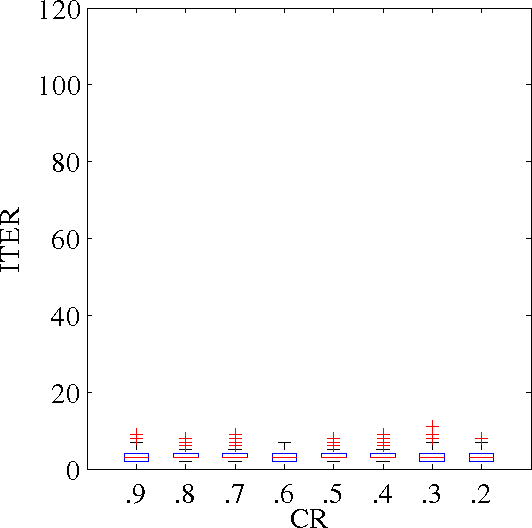}
    \label{subfig:gapITER}
}
\subfigure[]{
	\includegraphics[width = 3cm]{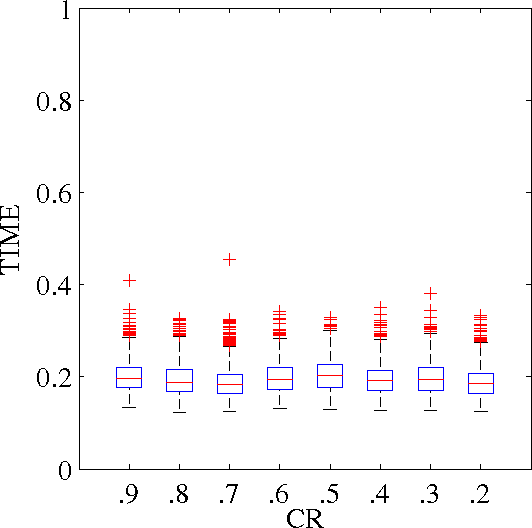}
    \label{subfig:gapTIME}
}
\caption[]{ Number of iterations (left) and processing times (right) until convergence of (\subref{subfig:sommpITER},\subref{subfig:sommpTIME}) SOMMP, (\subref{subfig:sgapITER},\subref{subfig:sgapTIME}) SGAP, (\subref{subfig:ommpITER},\subref{subfig:ommpTIME}) OMMP  and (\subref{subfig:gapITER},\subref{subfig:gapTIME}) GAP as function of CR. }
\label{fig:TIME}
\end{figure}

\subsection{Reconstruction Speed}

The processing times and number of iterations until convergence are presented in Fig. \ref{fig:TIME}. The (S)GAP requires only a few iterations to converge compared to (S)OMMP. This is because the co-support estimate in (S)GAP is merely used as a condition for the inversion problem and therefor does not have to be very close to the actual co-support. The support estimate in (S)OMMP, on the other hand, actually determines which elements in the dictionary are used for the signal estimate and therefor needs to be close to the actual support for the algorithm to converge to a solution.

Note that during each iteration 4 elements are added to the support in (S)OMMP, while in (S)GAP 10 elements are removed from the co-support. These numbers should be equal for a comparison of the number of iterations of both algorithms, though the goal of this paper is to compare their optimal reconstruction accuracy.

Despite the large difference in the number of iterations, SOMMP is only slightly slower than SGAP, though SGAP processing times vary less than those of SOMMP. This is because SOMMP solves a smaller inversion problem in each iteration, since it slowly builds a support starting from an empty one and only estimates coefficients corresponding to the support. SGAP, on the other hand, removes elements from an initally full co-support and uses this knowledge to solve a large inversion problem to find the best possible estimate of the original signal.

With decreasing CR, the total number of iterations of SOMMP decreases due to the fact that its stopping criterion depends on the norm of the residual, which reduces with each iteration. This residual, which is based on the original measurement, will be smaller to begin with and reduces to a sufficiently small value faster at higher CR. In SGAP, the number of iterations remains similar for each CR value.

The total processing time required to process both channels by GAP or OMMP is significantly higher than the time required by SGAP or SOMMP to process both channels at once. This indicates that it is indeed beneficial in terms of processing times to use an MC algorithm instead of an SC algorithm to process multiple channels. The reduction in time when using SGAP instead of GAP is also larger than the reduction obtained by using SOMMP instead of OMMP.

\begin{figure}[ht]
\centering
\subfigure[]{
	\includegraphics[width = 6cm]{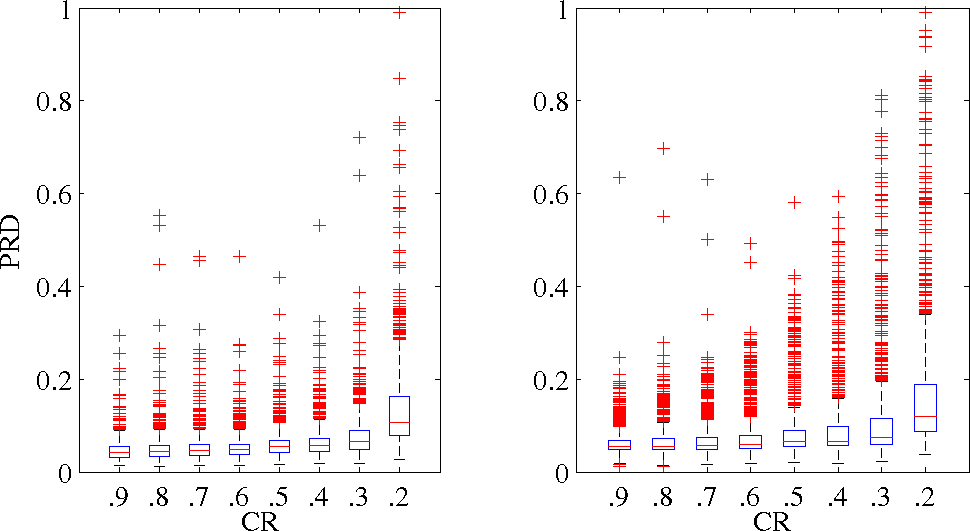}
    \label{subfig:sommpPRD}
}
\subfigure[]{
	\includegraphics[width = 6cm]{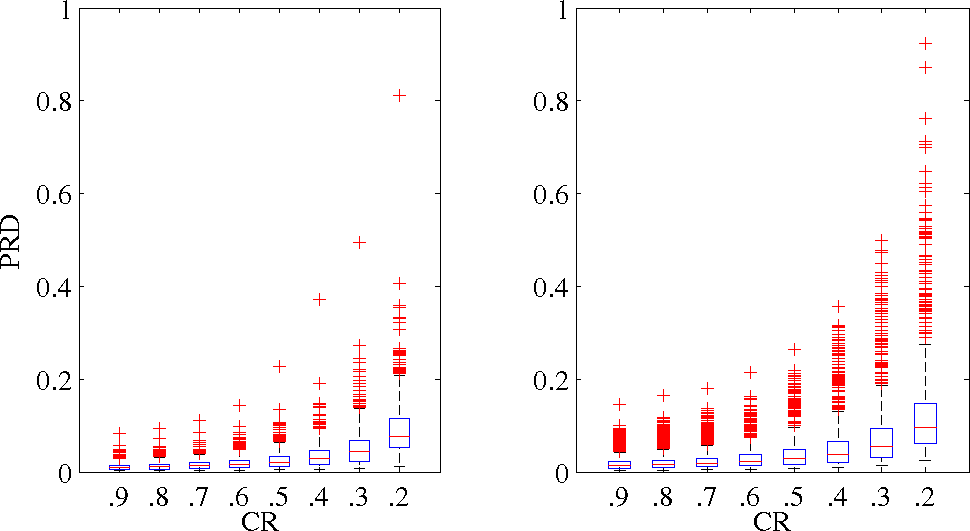}
    \label{subfig:sgapPRD}
}
\subfigure[]{
	\includegraphics[width = 6cm]{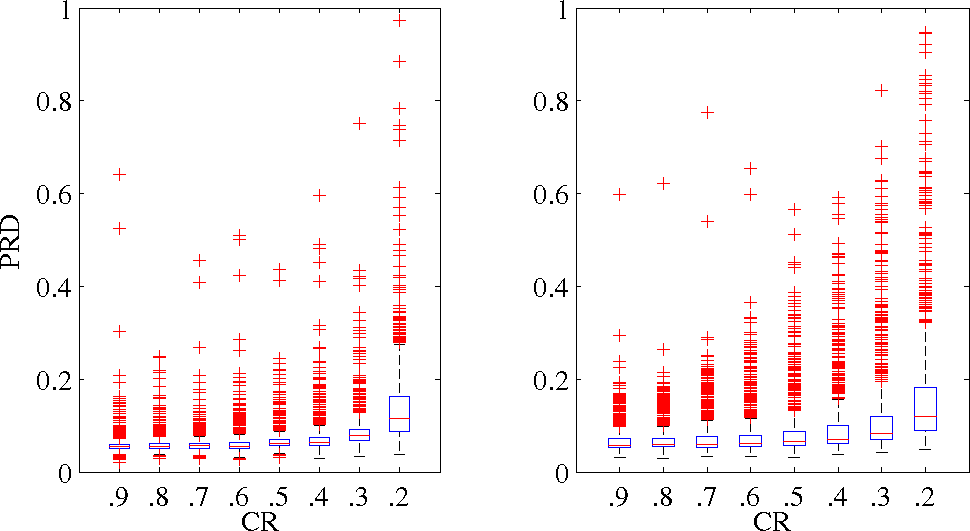}
    \label{subfig:ommpPRD}
}
\subfigure[]{
	\includegraphics[width = 6cm]{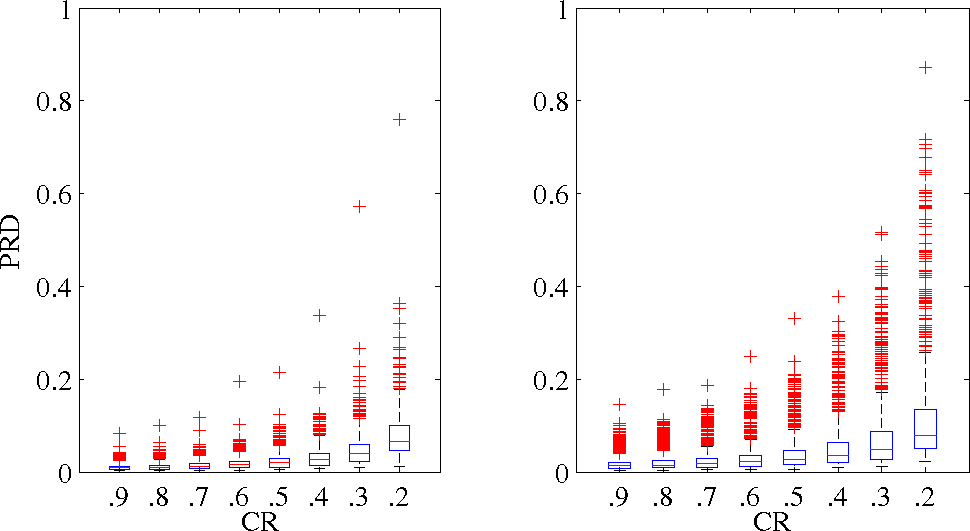}
    \label{subfig:gapPRD}
}
\caption[]{ PRD values of \subref{subfig:sommpPRD} SOMMP,  \subref{subfig:sgapPRD} SGAP, \subref{subfig:ommpPRD} OMMP and \subref{subfig:gapPRD} GAP as a function of CR for the 1st channel (left) and 2nd channel (right). }
\label{fig:PRD}
\end{figure}

\subsection{Reconstruction Accuracy}

PRD values of all algorithms are presented for both channels in Fig. \ref{fig:PRD}. The PRD values of the cosparse algorithms are far below those of the sparse algorithms at all CR values. As expected, the quality decreases with increasing CR values for all algorithms. The sparse algorithms also show more outliers compared to their cosparse counterparts.

Both MC algorithms show little difference in reconstruction accuracy, compared to their SC equivalents, though joint sparsity is expected to increase the accuracy on condition that all channels share the same support. 
In ECG, the support of the projections of different channels may be different because of the different electrodes positions. However, because each projection originates from the same source (the heart), the supports are similar enough to apply MC CS, but the benefits of using joint sparsity will be lost.

These results indicate that MC algorithms could be an interesting alternative to their SC versions, as long as the (co-)support of the different channels is similar. They also indicate that using cosparse algorithms in ECG applications is in fact advantageous to the use of their sparse counterparts.

\section{Conclusion}
\label{sec:conclusion}

In this paper, a cosparse algorithm for simultaneous reconstruction of MC ECG from CS measurements, was presented. It is shown that the algorithm reduces processing times, compared to the equivalent SC algorithm. Despite the fact that the mean processing time of SGAP is slightly higher than that of SOMMP, SGAP processing times vary less.


%
Simultaneous reconstruction of an MC ECG signal could also be achieved by running multiple GAP algorithms in parallel. However, it would require significantly more processing power, which is not always available.





\end{document}